\documentclass[preprint2]{aastex}

\def\plotfiddle#1#2#3#4#5#6#7{\centering \leavevmode
\vbox to#2{\rule{0pt}{#2}}
\includegraphics{#1}}

\def\lesssim{\mathrel{\hbox{\rlap{\hbox{\lower4pt\hbox{$\sim$}}}\hbox{$<$}}}}
\def\gtrsim{\mathrel{\hbox{\rlap{\hbox{\lower4pt\hbox{$\sim$}}}\hbox{$>$}}}}

\newcommand{\uvec}{\mbox{\boldmath $u$}}
\newcommand{\rvec}{\mbox{\boldmath $r$}}
\newcommand{\xvec}{\mbox{\boldmath $x$}}
\newcommand{\yvec}{\mbox{\boldmath $y$}}

\begin{document}

\title{On the Feasibility of Detecting Satellites of Extrasolar Planets 
        via Microlensing}

\author{Cheongho Han}
\affil{Department of Physics, Chungbuk National University,
       Chongju 361-763, Korea}
\email{E-mails: cheongho@astroph.chungbuk.ac.kr} 

\author{Wonyong Han}
\affil{Korea Astronomy Observatory, Taejon 305-348, Korea}
\email{whan@kao.re.kr}

\begin{abstract}
Although many
methods of detecting extra-solar planets have been proposed and 
successful implementation of some of these methods enabled a rapidly 
increasing number of exoplanet detections, little has been discussed 
about the method of detecting satellites around exoplanets.  In this 
paper, we test the feasibility of detecting satellites of exoplanets 
via microlensing. For this purpose, we investigate the effect of 
satellites in the magnification pattern near the region of the 
planet-induced perturbations by performing realistic simulations of 
Galactic bulge microlensing events. From this investigation, we find 
that although satellites can often cause alterations of magnification 
patterns, detecting satellite signals in  lensing light curves will be 
very difficult because the signals are seriously smeared out by the 
severe finite source effect even for events involved with source stars 
with small angular radii.
\end{abstract}

\section{Introduction}

Various methods have been proposed to search for extrasolar planets 
(exoplanets).  These methods include the pulsar timing analysis, direct 
imaging, accurate measurement of astrometric displacements, radial 
velocity measurement, planetary transit, and gravitational microlensing 
[see the review of \citet{perryman00}].  Since the first detection of an 
exoplanet around the pulsar PSR 1257+12 \citep{wolszczan92}, nearly 100 
exoplanets have been identified (http://exoplanets.org), mostly by the 
radial velocity method \citep{mayor95}.

However, little has been discussed about the method of detecting 
satellites around exoplanets.  This is mainly because it is thought to 
be too premature to detect satellites given the difficulties of detecting 
exoplanets.  Currently, the only promising technique proposed to detect 
satellites is the transit method, where satellites are detected either by 
direct satellite transit or through perturbations in the transit timing of 
the satellite-hosting planet \citep{sartoretti99}.

In this paper, we investigate the feasibility of detecting satellites 
of exoplanets via microlensing.  Detection of a low-mass companion by 
using microlensing is possible because the companion can induce noticeable 
anomalies in the resulting lensing light curves \citep{mao91, gould92}.
The microlensing method has an important advantage in detecting very 
low-mass companions over other methods because the strength of the 
companion's signal depends weakly on the companion/primary mass ratio 
although the duration of the signal becomes shorter with the decrease of 
the mass ratio.  Then, if lensing events are monitored with a high enough 
frequency, it may be possible to detect not only planets but also their 
satellites.  Such a frequent lensing monitoring program in space was 
recently proposed by \citet{bennett00}.

The paper is organized as follows.  In \S\ 2, we describe the microlensing 
basics of multiple-lens systems.  In \S\ 3, we investigate the feasibility 
of satellite detections by carrying out realistic simulations of Galactic 
bulge microlensing events caused by an example lens system having a planet 
and a satellite.  We summarize the results and conclude in \S\ 4.

\section{Multiple-lens Systems}
To describe the 
lensing behaviors of events caused by a lens system composed of a primary
with a planet and subordinate satellites, it is required to have the 
formalism of multiple-lens systems.  If a source located at $\rvec_{\rm S}
(\xi, \eta)$ on the projected plane of the sky is lensed by a lens system 
composed of $N$-point masses, where the individual components' masses and 
locations are $m_{i}$ and $\rvec_{{\rm L},i}$, the positions of the resulting 
images, $\rvec_{\rm I}$, are obtained by solving the lens equation, which 
is expressed by
\begin{equation}
\rvec_{\rm S} = \rvec_{\rm I} - \theta_{\rm E}^{2} 
\sum_{i=1}^{N} {m_i\over m} 
{\rvec_{\rm I}-\rvec_{{\rm L},i} \over 
\left\vert\rvec_{\rm I}-\rvec_{{\rm L},i}\right\vert^2}.
\end{equation}
Here $\theta_{\rm E}$ represents the angular Einstein ring radius, which is 
related to the total mass of the lens system, $m=\sum_{i}^{N} m_{i}$, and 
the distances to the lens, $D_{\rm OL}$, and the source, $D_{\rm OS}$, by
\begin{equation}
\theta_{\rm E}=\sqrt{{4Gm\over c^2}} 
               \left( {1\over D_{\rm OL}}-{1\over D_{\rm OS}}\right)^{1/2}.
\end{equation}
The lensing process conserves the surface brightness of the source.  Then 
the magnification of each image equals to the surface area ratio between 
the image and the unmagnified source and mathematically its value corresponds 
to the inverse of the Jacobian of the lens equation evaluated at the image 
position $\rvec_{{\rm I},j}$, i.e.\
\begin{equation}
A_{j}=\left( {1\over \left\vert\det J\right\vert} \right)_{\rvec_{\rm I}=
\rvec_{{\rm I},j}} ;
\qquad 
\det J= \left\vert {\partial \rvec_{\rm S} 
\over \partial \rvec_{\rm I}} \right\vert.
\end{equation}
Although the magnifications of the individual images cannot be measured
due to the small separations between the images, one can measure the total 
magnification, i.e.\ $A=\sum_{j}^{N_{\rm I}} A_{j}$, where $N_{\rm I}$ is 
the total number of images.  Note that to find the image positions and 
the magnification, it is required to invert the lens equation.

For a single point-mass lens ($N=1$), the lens equation can be easily 
inverted.  Solving the equation yields  two solutions of image positions 
and the total magnification is expressed in a simple analytical form of
\begin{equation}
A = {u^2+2 \over u \sqrt{u^2+4}},
\end{equation}
where $\uvec=(\rvec_{\rm S}-\rvec_{\rm L})/\theta_{\rm E}$ is the 
dimensionless lens-source separation vector normalized by $\theta_{\rm E}$.  
For a rectilinear lens-source transverse motion, the separation vector is 
related to the single lensing parameters by
\begin{equation}
\uvec = 
\left({t-t_{0}\over t_{\rm E}}\right)\ \hat{\xvec}\ +\ \beta\ \hat{\yvec},
\end{equation}
where $t_{\rm E}$ represents the time required for the source to transit
$\theta_{\rm E}$ (Einstein time scale), $\beta$ is the closest lens-source
separation in units of $\theta_{\rm E}$ (impact parameter), $t_0$ is the
time of the maximum magnification, and the unit vectors $\hat{\xvec}$ and 
$\hat{\yvec}$ are parallel with and normal to the direction of the relative 
lens-source transverse motion, respectively.  The light curve of a single 
point-mass lens event is characterized by its smooth and symmetric shape 
\citep{paczynski86}.

If the lens system has additional components ($N\geq 2$), the lens equation
cannot be algebraically inverted.  However, the 
lens equation can be expressed as a polynomial in $\rvec_{\rm I}$ and the 
positions of the individual images are obtained by numerically solving the 
polynomial \citep{witt90}.  If the lens system is composed of two lenses 
(e.g.\ the primary and the plant), the lens equation is equivalent to a 
fifth-order polynomial in $\rvec_{\rm I}$ and there exist three or five 
solutions of the image positions depending on the source location with 
respect to the lens components.  The main new feature of the multiple-lens 
system is the caustics, which represent the set of points in the source 
plane where the magnification of a point source becomes infinity, i.e.\ 
$\left\vert\det J \right\vert=0$.  Hence a significant planet-induced 
deviation in the lensing light curve occurs when the source approaches the 
region around the caustic although the duration of the deviation is short 
due to the small mass ratio between the planet and the primary, $q_{\rm p}$.  
The size of the caustic, and thus the probability of planet detections, also 
depends on the primary-planet separation and is maximized when the separation 
(normalized by $\theta_{\rm E}$) is in the range of $0.6 \lesssim d_{\rm p} 
\lesssim 1.6$ (lensing zone).

As the number of lens components increases, solving the lens equation 
becomes nontrivial because the order of the polynomial increases by $N^2+1$.  
One other method commonly used to obtain the magnification patterns of 
multiple-lens systems is the inverse ray-shooting technique \citep{schneider86, 
kayser86, wambsganss97}.  In this method, a large number of light rays are 
uniformly shot backwards from the observer plane through the lens plane 
and then collected (binned) in the source plane.  Then, the magnification 
pattern of the lens system is obtained by the ratio of the surface brightness 
(i.e., the number of rays per unit area) in the source plane to that in 
the observer plane.  Once the magnification pattern is constructed, the 
light curve resulting from a particular source trajectory corresponds to 
the one-dimensional cut through the constructed magnification pattern.  
The advantage of using the ray-shooting method is that it allows one to 
study the lensing behavior regardless of the number of lens components.
The disadvantage is that it requires a large computation time for the 
construction of the detailed magnification pattern.  We attempted to 
investigate the lensing behaviors of multiple lens systems by solving the 
polynomial but we found that for the lens system of our interest for which 
the mass ratio between the least (satellite) and the most massive (primary) 
components is smaller than $\sim 10^{-5}$, the numerical noise in the 
polynoimial coefficients due to the limited computer precision ($\sim 10^{-15}$) 
causes serious inaccuracy in solving the polynomial.  We, therefore, use the 
ray-shooting method despite the requirement of large computation time.

\section{Realistic Simulations}

To examine the feasibility of detecting satellites of exoplanets,
we carry out realistic simulations of Galactic bulge microlensing events 
caused by a lens system having a planet and a satellite.  The primary of 
the tested lens system is assumed to have a mass of $0.3\ M_\odot$ by 
adopting that of a late-type main-sequence star, which is believed to be 
the most common type of lenses for events detected towards the Galactic 
bulge \citep{alcock00}.  For the planet and the satellite, we test an 
Earth-mass planet and a Moon-mass satellite.  Then, the mass ratios of 
the planet and the satellite with respect to the primary are $q_{\rm p}
=10^{-5}$ and $q_{\rm s}=1.2\times 10^{-7}$, respectively.  The planet is 
assumed to be separated by $d_{\rm p}=1.3$ from the primary.  To investigate 
the dependence of the magnification pattern on the satellite position with 
respect to the planet, we test four cases of satellite locations with 
different combinations of the planet-satellite separation, $d_{\rm s}$ 
(normalized by the Einstein ring radius of the planet $\theta_{\rm E,p}=
\sqrt{q_{\rm p}}\theta_{\rm E}$), and the orientation angle, $\alpha$ 
(measured from the primary-planet axis).

For a low-mass companion, the source size can have a significant effect 
on the shape of the companion-induced anomalies in lensing light curves 
\citep{bennett96}.  For the construction of light curves, we, therefore, 
test three different types of source stars with angular radii of $\rho_\star
=\theta_\star/\theta_{\rm E}=1.5\times 10^{-3}$, $1.1\times 10^{-3}$, and 
$0.85\times 10^{-3}$.  These values correspond to the source star radii of 
F0, G0, and K0 main-sequence stars located at $D_{\rm OS}=8.5$ kpc for an 
event with a lens located at the half way point between the observer and the 
source, i.e.\ $D_{\rm OL}/D_{\rm OS}=0.5$.  We note that although a M-type 
star has a smaller source radius ($\rho_\star= 0.6\times 10^{-3}$ for M0 
main sequence), we do not consider the case because the star will be too 
faint to be observed ($I \geq 25.6$ even without extinction).
For comparison, we also present the light curves corresponding to a source 
with $\rho_\star=3.0\times 10^{-5}$, which is equivalent to the radius of 
a white dwarf.  Although these events would be rare, there is at least 
some finite probability that a caustic crossing of a white dwarf would occur.
In addition, although the brightness of a hot white dwarf would be faint
($V\gtrsim 25$), this is not much fainter than a typical K dwarf.
Therefore, instead of presenting light curves of events involved with a 
hypothetical source star having an arbitrary radius, we present the light 
curves of events involved with a hot white dwarf.

\begin{figure*}
\plotfiddle{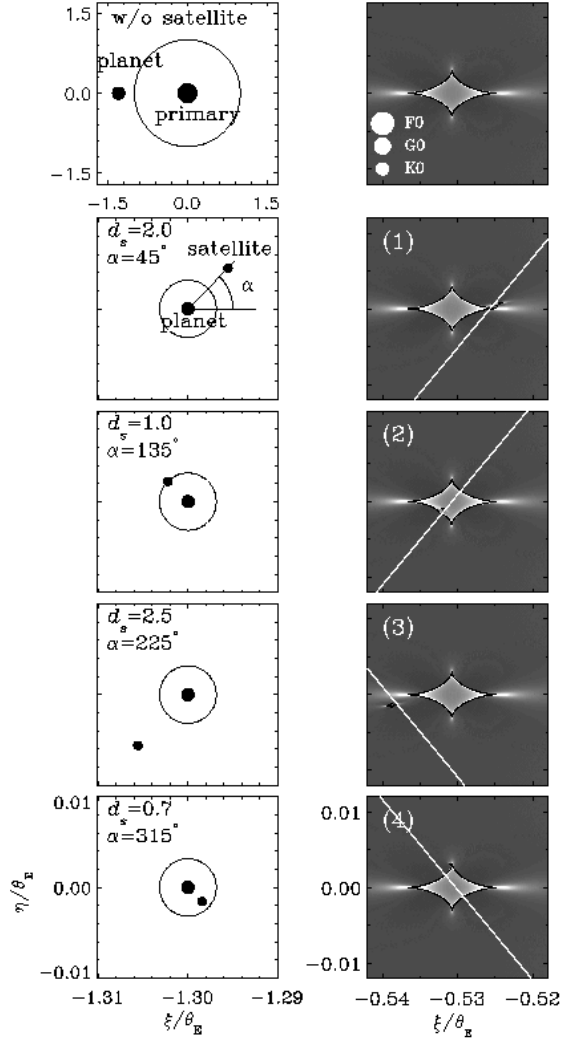}{0.0cm}{0}{50}{50}{-150}{-380}
\vskip13.5cm
\caption{
Magnification patterns of a lens system composed of a primary, a planet,
and a subordinate satellite (gray-scale maps in the right panels) along
with the corresponding geometries of the lens systems (left panels).
For both the magnification maps and the lens system geometries, the
coordinates are centered at the center of mass of the lens system and
all lengths are normalized by $\theta_{\rm E}$.  The circle in the upper
left panel and those in the other left panels represent the Einstein
rings of the primary and the planet (with a radius $\theta_{\rm E, p} = 
\sqrt{q_{\rm p}} \theta_{\rm E}$), respectively.  The planet is separated
by $d_{\rm p}=1.3$ from the primary.  The planet/primary and
satellite/primary mass ratios are $q_{\rm p}=10^{-5}$ and $q_{\rm s}=1.2 
\times 10^{-7}$, which correspond to a Earth-mass planet and a Moon-mass
satellite around a $0.3 M_\odot$ star, respectively.  The labels in each
of the left panel represent the planet-satellite separation, $d_{\rm s}$
(normalized by $\theta_{\rm E,p}$), and the orientation angle, $\alpha$
(measured from the primary-planet axis).  The solid curve in the
magnification map represents the caustics.  For comparison, we present
the magnification pattern unperturbed by the satellite in the upper
right panel.  The three white dots in the upper right panel represent
the source sizes (corresponding to those of F0, G0, and K0 main-sequence
stars) which are used to construct the light curves in Fig.\ 2.  The
white lines in the right panels represent the source trajectories of
the events, whose resulting light curves are presented in Fig.\ 2.
}
\end{figure*}

\begin{figure}
\plotfiddle{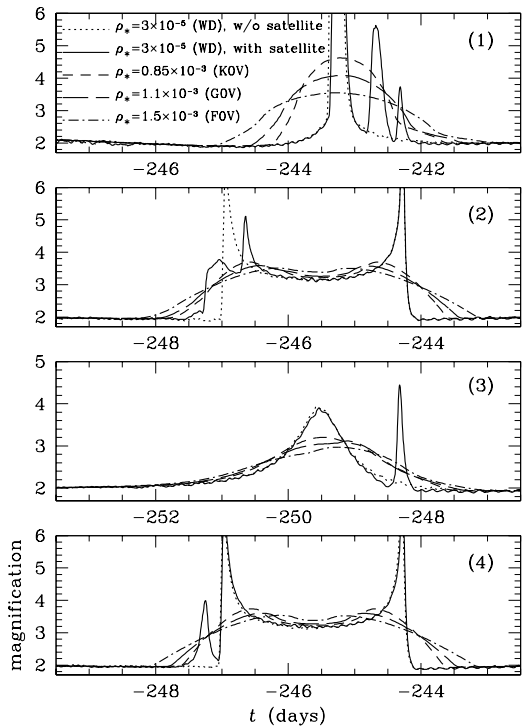}{0.0cm}{0}{50}{50}{-150}{-288}
\vskip7.4cm
\caption{Variation of the lensing light curve anomalies induced by an Earth-mass
planet having a Moon-mass satellite.  In each panel, we present three
different light curves involved with source stars having different
angular radii of $\rho_\star=\theta_\star/\theta_{\rm E}=0.85\times 10^{-3}$
(short-dashed curve), $1.1\times 10^{-3}$ (long-dashed curve), and
$1.5\times 10^{-3}$ (dot-dashed curve), which correspond to the source
radii of K0, G0, and F0 main-sequence stars for a Galactic bulge event
caused by a lens with a total mass $m=0.3M_\odot$ and located at
$D_{\rm OL}/D_{\rm OS}=0.5$.  For comparison, we also present the light
curves corresponding to the source of a white dwarf with $\rho_\star=3.0
\times 10^{-5}$, which are expected with (solid curve) and without the
satellite (dotted curve).  The source trajectories responsible for the
light curves are marked on the magnification maps in Fig.\ 1,  where the
corresponding panels of the source trajectories and the light curves are
marked by the same panel number.  The time is presented in days assuming
that the event has an Einstein timescale of $t_{\rm E}=30$ days.  The
reference of the time is arbitrarily set.}
\end{figure}

In Figure 1, we present the magnification patterns (gray-scale maps in 
the right panels) around the regions of deviations induced by the planet 
along with the corresponding geometries of the lens systems (left panels).  
In Figure 2, we also present the light curves of events resulting from the 
source trajectories marked in the corresponding panels in Fig.\ 1.  From 
the simulations, we find the following results.
\begin{enumerate}
\item
If the satellite is located within the lensing zone of the planet, i.e.\ 
$0.6 \lesssim d_{\rm s} \lesssim 1.6$\footnote{For a system of the Earth 
and the Moon located at $D_{\rm OL}/D_{\rm OS}\sim 0.5$, the Earth-Moon 
separation is about 3 times of the Einstein ring radius of an Earth-mass 
planet in the foreground of the Galactic bulge.}, the planet-induced 
caustic shape and the magnification pattern around the caustics are altered 
by the satellite due to the interference between the anomalies induced by 
the planet and the satellite [see panel (2) of Fig.\ 1].
\item
If the satellite-planet separation is larger than the upper limit of the 
lensing zone of the planet, i.e.\ $d_{\rm s}\gtrsim 1.6$, the interference 
becomes negligible.  Then, the resulting magnification pattern is well 
represented by the superposition of those of the two binary systems where 
the planet-primary and the satellite-primary pairs act as independent lens 
systems \citep{bozza99, han01, han02, rattenbury02}.  Although the satellite 
is located beyond the planet's lensing zone, the satellite-planet separation 
is generally much smaller than the separation between the primary and the 
planet, i.e.\ $d_{\rm s}\theta_{\rm E,p}\ll d_{\rm p}\theta_{\rm E}$.  
Therefore, the additional deviations induced by the satellite is located near 
the region of planet-induced deviations.
\item
If the satellite-planet separation is smaller than the lower limit of the 
lensing zone of the planet, i.e.\ $d_{\rm s}\lesssim 0.6$, the planet-satellite 
pair behavior as if they are a single lens component with a mass equal to the 
combined one of the planet and the satellite.  Since satellites are generally 
much less massive than their planets, i.e.\ $q_{\rm p}+q_{\rm s}\sim q_{\rm p}$, 
and thus the lensing behavior in this case is hardly affected by the presence 
of the satellite.
\item
The size (area) of the satellite-induced perturbation region is comparable or 
smaller than the size of source stars that can be monitorred from followup 
lensing observations.  As a result, the detailed structure in the lensing light 
curves is seriously smeared out due to the finite source effect.  We find that 
the finite source effect is so severe that the satellite signals in the light 
curves of all tested events are completely washed out even for events involved 
with K0 source stars.
\end{enumerate}

\section{Summary and Conclusion}
We have tested the feasibility of detecting satellites by using microlensing. 
For this purpose, we have investigated the effect of satellites on the 
magnification pattern near the region of planet-induced perturbations by 
carrying out realistic simulations of Galactic bulge microlensing events.
From this investigation, we find that although satellites can often affect 
the maginification patterns, detecting satellite signals in the lensing light 
curves will be very difficult because the signals are seriously smeared out by 
the severe finite source effect.

\bigskip
\bigskip
We would like to thank J.\ H.\ An for making useful comments about the work.  
This work was supported by a grant (2001-DS0074) from Korea Research 
Foundation (KRF).

{}

\end{document}